\documentclass[12pt]{nature}
\usepackage{graphicx}
\usepackage{caption}

\title{Collisions of matter-wave solitons}
\author{Jason H.V. Nguyen$^{1}$, Paul Dyke$^{2}$, De Luo$^{1}$, Boris A.
Malomed$^{3}$ \& Randall G. Hulet$^{1\ast}$}

\begin{document}

\maketitle

\begin{affiliations}
 \item Department of Physics and Astronomy, Rice University, Houston, TX 77005
 \item Centre of Quantum and Optical Science, Swinburne University of Technology,
 Melbourne 3122, Australia
 \item Department of Physical Electronics, School of Electrical Engineering,
 Faculty of Engineering, Tel Aviv University, Tel Aviv 69978, Israel
\end{affiliations}

\begin{abstract}
Solitons are localised wave disturbances that propagate without changing
shape, a result of a nonlinear interaction which compensates for wave packet
dispersion. Individual solitons may collide, but a defining feature is that
they pass through one another and emerge from the collision unaltered in
shape, amplitude, or velocity, but with a new trajectory reflecting a discontinuous jump.  This remarkable property is mathematically a
consequence of the underlying integrability of the one-dimensional (1D)
equations, such as the nonlinear Schr\"odinger equation, that describe
solitons in a variety of wave contexts, including matter-waves\cite{Zabusky65, Zakharov72}. Here we explore the nature of soliton collisions
using Bose-Einstein condensates of atoms with attractive interactions
confined to a quasi-one-dimensional waveguide. We show by real-time imaging
that a collision between solitons is a complex event that differs markedly
depending on the relative phase between the solitons. %Yet, they emerge from the collision unaltered in shape or amplitude, but with a new trajectory reflecting a discontinuous jump. 
By controlling the strength of the
nonlinearity we shed new light on these fundamental features of soliton
collisional dynamics, and explore the implications of collisions in the proximity of the crossover between one and three dimensions where the loss of integrability may precipitate catastrophic collapse.
\end{abstract}

The name ``soliton'' is meant to convey the particle-like qualities of
localised and non-dispersing wave packets, and is often reserved for those
solitary waves that pass through one another without changing form. The wave packets studied here are not true solitons due to the presence of a harmonic confining potential which breaks integrability.  Furthermore, as quasi-$1$D objects, they reside near the border between $1$D and $3$D, where integrability is also broken.  Nonetheless, the proximity to this border does not necessarily affect the soliton dynamics over experimentally relevant time scales.   We will thus use the term soliton more generally to
refer to non-dispersing wave packets created by a nonlinearity. While the
propagation of individual solitons is now well-understood, having been
extensively studied both experimentally and theoretically\cite{Chen12},
their interactions with each other have been much less explored. While true solitons pass through one another, they nonetheless exhibit an
effective interaction produced by interference of the two wave packets, with
a force falling off exponentially with separation\cite{Gordon83}. The sign
and magnitude of the interaction depends on the relative phase, as was first
demonstrated experimentally for optical solitons in both the time\cite%
{Mitschke87} and spatial\cite{Aitchison91} domains.

Studies of matter-wave solitons have mainly examined properties of single solitons%
\cite{Khaykovich02, Strecker02, Marchant13, Medley14}, with the study of
soliton interactions limited to those occurring in soliton trains\cite%
{Strecker02, Kawaja02} and collisions between multiple solitons resulting
from a quench\cite{Cornish06}. We provide further insight into the collisional dynamics of matter-wave 
solitons through the controlled formation of soliton pairs, and explicitly demonstrate 
that the discontinuous jump observed in soliton collisions\cite{Islam91} is a general
property of the nonlinear interaction.

\begin{figure}
\centering
\includegraphics{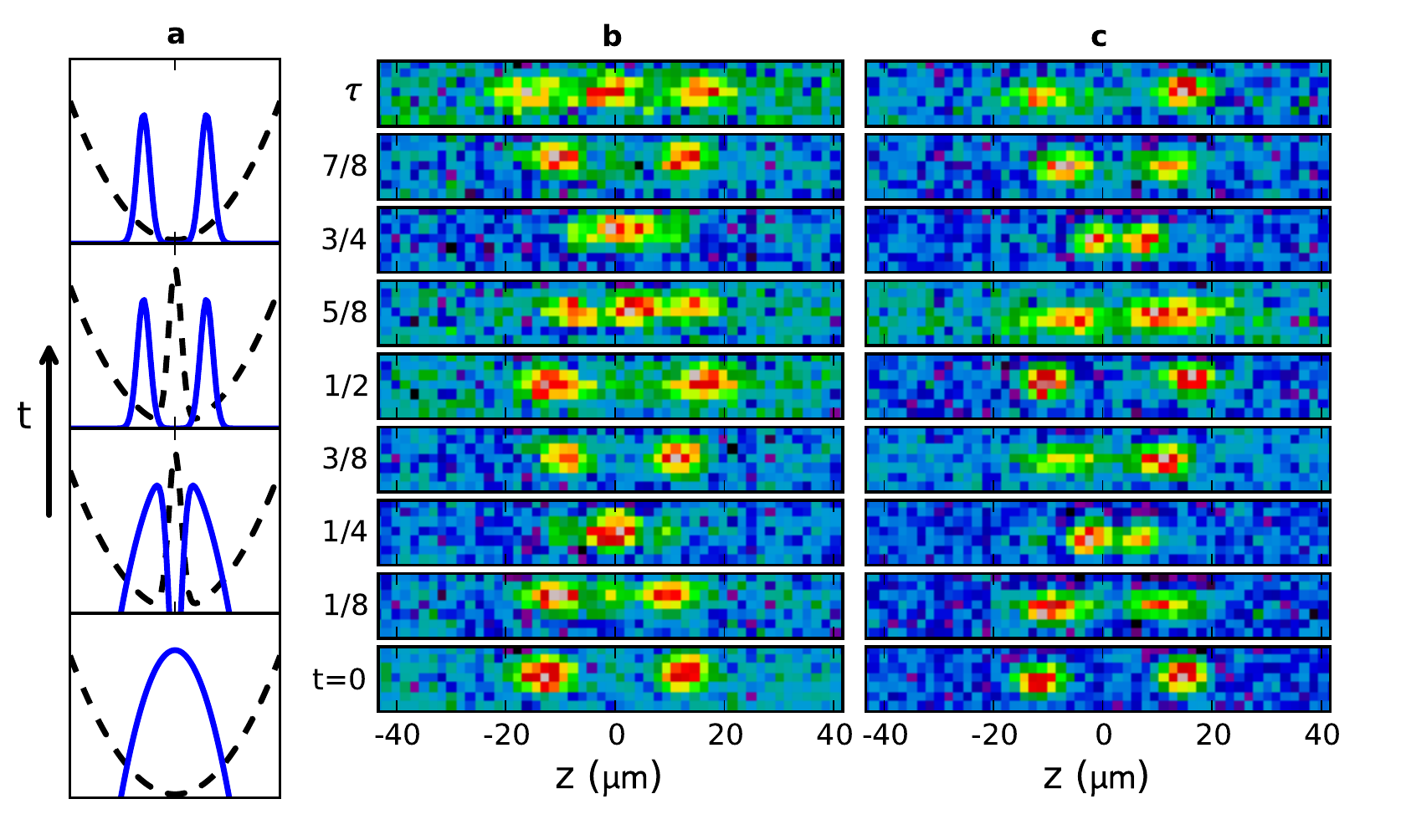}
\caption{\textbf{Schematic of the experiment and images of phase-dependent
collisions.} \textbf{a}, Schematic showing the process of soliton-pair
formation. Beginning with the bottom frame, the potential is shown as a
black-dashed line with a condensate density profile shown in solid blue.
After forming a condensate, the barrier is turned on to split the condensate
in two. The scattering length is ramped from ${a=+140a_{0}}$ to ${%
a=-0.57a_{0}}$ and pairs of solitons are formed. The barrier is quickly
turned off, and the solitons move towards the centre of the trap. \textbf{b},
Time evolution of a soliton pair (${N/ N_{c}=-0.53}$) after the barrier is
turned off. Solitons are accelerated towards the centre of the trap and
collide at a quarter-period (${\protect\tau=2\protect\pi / \protect\omega%
_{z} = 32\ \mathrm{ms}}$). The density peak appearing at the centre-of-mass
indicates that this is an in-phase ($\Delta \protect\phi \approx 0$)
collision. \textbf{c}, Similar to \textbf{b}, except the density node appearing
at the centre-of-mass indicates an out-of-phase ($\Delta \protect\phi %
\approx \protect\pi$) collision.}
\label{fig:fig1}
\end{figure}

Our methods for producing a degenerate gas of ${^{7}\mathrm{Li}}$ atoms have been described previously\cite{Dries10} and are summarized in the Methods section. A Bose-Einstein condensate of
atoms in the ${|\mathrm{F}=1,\mathrm{m_{F}}=1\rangle}$ state is formed by
evaporative cooling at a scattering length of ${a=+140a_{0}}$ and is
confined in a cylindrically symmetric harmonic trap with radial and axial
oscillation frequencies of ${\omega_{r}/2 \pi=254\ \mathrm{Hz}}$ and ${%
\omega_{z}/2 \pi=31\ \mathrm{Hz}}$, respectively. After forming the
condensate, a cylindrically-focussed blue-detuned Gaussian laser beam
directed perpendicular to the long axis of the confining potential is used
to cut the condensate in half, and acts as a barrier between the two
condensates (Fig. \ref{fig:fig1}a). The scattering length is then
adiabatically ramped from ${a=+140a_{0}}$ to ${a=-0.57a_{0}}$ via the
broadly-tunable Feshbach resonance of the ${|\mathrm{F}=1,\mathrm{m_{F}}%
=1\rangle}$ state\cite{Pollack09} to form a pair of solitons with a
centre-to-centre separation of ${26\ \mathrm{\mu m}}$ and near equal
amplitude ($N\approx 28,\!000$ atoms / soliton). Once the pair is formed, the
barrier is quickly (${t<60\ \mathrm{ns}}$) turned off. Thus, the solitons
suddenly find themselves at the classical turning points of the harmonic
trap and begin to accelerate towards the centre. We confirm that these wave
packets are nondispersive by observing the absence of expansion when the
axial confinement frequency is suddenly reduced, while a wave packet with
a small, repulsive scattering length rapidly expands (see Supplementary Fig.~$1$%
).

We find that the two solitons interact with a randomly distributed relative phase over
different experimental runs, so we use a minimally-destructive phase
contrast imaging method, polarisation phase-contrast imaging\cite{Bradley97}%
, to obtain multiple images of the soliton pair as they oscillate and
collide in the harmonic trap. This imaging technique plays a key role for
observing and interpreting the collisional dynamics since it allows us to
take multiple images within a single realisation of the experiment (see
Methods).

We infer the relative phase difference through comparison with numerical
simulations of the $1$D and $3$D Gross-Pitaevskii equations (GPEs). Figures~\ref{fig:fig1}b and \ref{fig:fig1}c show two experimental
realisations in which the relative phase difference is ${\Delta \phi \approx
0}$ and ${\Delta \phi \approx \pi}$, respectively. Images are taken every
one eighth of a trap period (${\tau=32\ \mathrm{ms}}$). Figures~\ref%
{fig:fig1}b and \ref{fig:fig1}c show trajectories over one complete period,
corresponding to two collisions. For ${\Delta\phi \approx 0}$, a clear anti-node is
observed during the collision at the centre, giving the appearance of an
attractive interaction, while in the ${\Delta\phi \approx \pi}$ case, interference
results in a central node and the interaction between solitons is
effectively repulsive.

\begin{figure}
\centering
\includegraphics{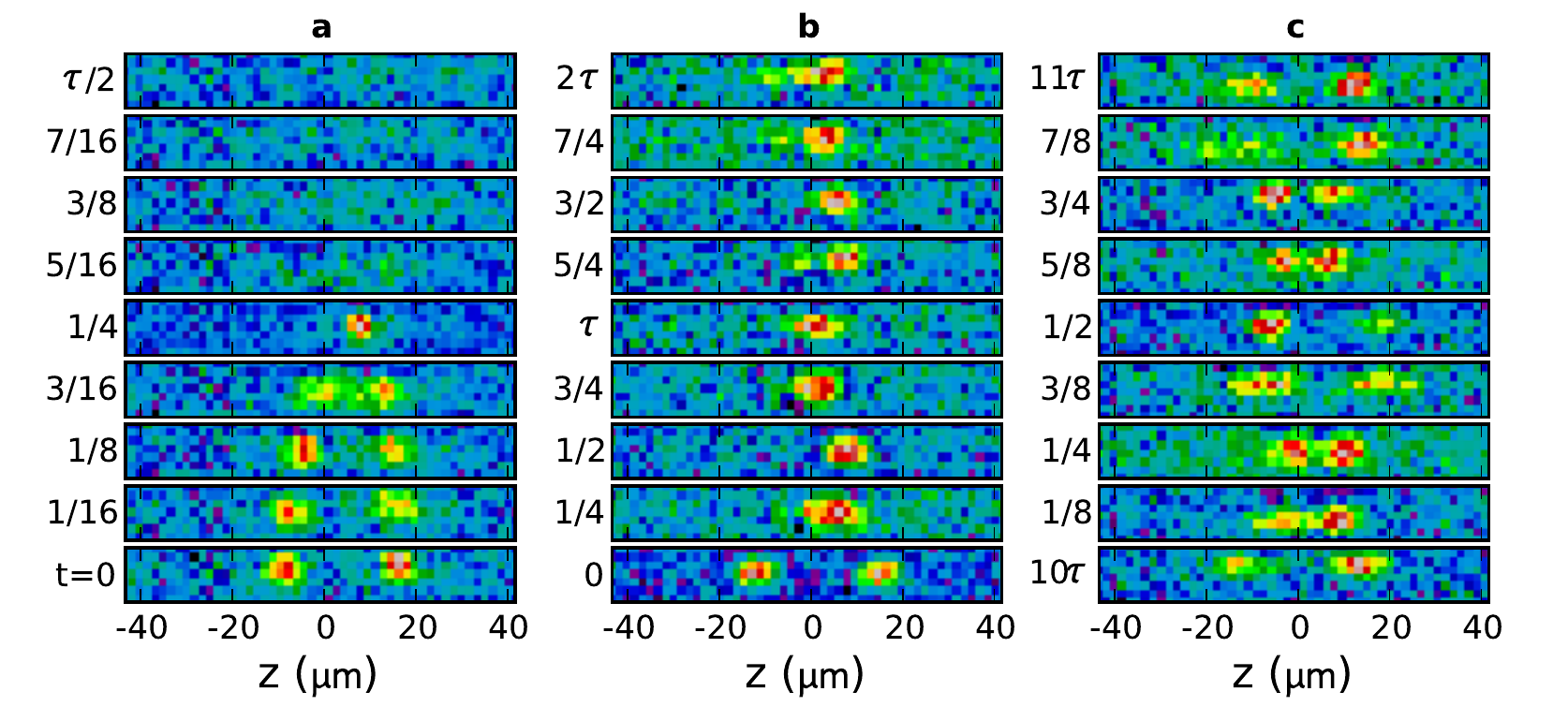}
\caption{\textbf{Phase-dependent collisional dynamics.} \textbf{a}, A
collision between two solitons (${N/ N_{c}=-0.53}$) resulting in collapse.
During the collision, the density exceeds a critical value and becomes
unstable against collapse. No remaining atoms are observed. \textbf{b}, A
collision between two solitons (${N/ N_{c}=-0.53}$) resulting in a merger.
The remaining atom number after the collision is the same as that of a
single soliton before the collision. \textbf{c}, Out-of-phase collisions
between two solitons after allowing them to oscillate for ten trap periods.}
\label{fig:fig2}
\end{figure}

The quasi-one-dimensional nature of our system coupled with the ability to
form soliton pairs with a strong nonlinearity allows us to observe the rich
dynamics inherent in a system at the edge of integrability. The strength of
the nonlinearity is parametrised by $N/N_{c}$, where $N_{c}=0.67a_{r}/a$ is
the critical atom number, ${a_{r}=\sqrt{\hbar /m\omega _{r}}}$, and m is the atomic mass. A
soliton is unstable to collapse for $N>|N_{c}|$\cite{Gammal01}. Although
collapse is relevant only for attractive interactions, we also use $N/N_{c}$
to parametrise the strength of the nonlinearity for repulsive condensates.
For values of ${N/N_{c}=-0.53}$, we observe that in-phase collisions ($%
\Delta \phi \approx 0$) sometimes result in annihilation (Fig.~\ref{fig:fig2}%
a), or fusion of the soliton pair (Fig.~\ref{fig:fig2}b), although more
typically we observe partial collapses in which the atom number and the
oscillation amplitude are reduced after multiple collisions. These effects
can be understood as the result of density-dependent inelastic collisions in
which the system becomes effectively three-dimensional\cite{Baizakov04, Khaykovich06, 
Parker08}. Similar effects have been observed in nonlinear optics\cite%
{Krolikowski98}. We find from the GPE simulations that collisions with ${%
\Delta \phi =0}$ and ${N/N_{c}<-0.5}$ are unstable to collapse. The
observation that collisions with $\Delta \phi \approx 0$ do not always lead
to collapse (e.g. Fig.~\ref{fig:fig1}b), is consistent with the shot-to-shot
variation in $N$ of $\sim\!20\%$ (see Methods). For the same nonlinearity,
out-of-phase collisions ($\Delta \phi \approx \pi $) are extremely robust
against collapse and survive many oscillations in the trap, as predicted
theoretically\cite{Baizakov04, Parker08, Billam11}. Although on the edge of
integrability, we have observed solitons with ${N/N_{c}=-0.53}$ and ${\Delta
\phi =\pi }$ to survive more than $20$ collisions (Fig.~\ref{fig:fig2}c).

The defining property of solitons passing through one another without change
of shape, amplitude, or speed seems to be at odds with the observations
presented in Fig.~\ref{fig:fig1}c, where solitons with $\Delta \phi =\pi $
apparently reflect from one another. This apparent paradox is resolved by
noting that the effective interaction is a wave phenomenon\cite{Gordon83},
where interference gives the appearance of reflection, when in fact, the
solitons do pass through one another. We experimentally demonstrate this by forming pairs of
solitons with unequal atom numbers by removing atoms from one side using a
short duration, near resonant pulse of light before ramping the field to
form solitons. This allows us to identify, or tag, a particular soliton and
to follow its trajectory before and after the collision. In Fig.~\ref%
{fig:fig3} we show one such realisation in which a soliton pair was formed
with a ${2\!\!:\!\!1}$ ratio in atom number. While a minimum does appear
between the solitons during the collision, as expected for an effectively
repulsive interaction, the trajectories show that they do pass through one
another.  The experiment does not rule out the possibility that the solitons reflect while exchanging particles during the collision.  The $1$D GPE simulations, however, demonstrate that particle exchange is a relatively small effect for the large collisional velocity in our experiment, in agreement with previous theoretical studies\cite{Khaykovich06, 
Parker08}.  

A close inspection of the oscillations shown in Fig.~\ref{fig:fig2}c reveals
that the solitons oscillate at a higher frequency than the usual dipole
frequency, $\omega_{z}$. This increased oscillation frequency is a
consequence of a jump in the phase of the trajectories of the colliding,
harmonically confined, solitons. Without axial confinement, the phase jump
manifests as a sudden change in position relative to the original
trajectory, as shown in the simulation of Fig. \ref{fig:fig4}a. Although the
phase of the trajectory is modified by the collision, the speed of the
soliton following the collision is not. The jump is a consequence of the nonlinearity of the system\cite{Zakharov72}, and was first observed with optical solitons\cite{Islam91}.

We studied this effect by measuring the oscillation frequency of pairs of
solitons for different strengths and sign of the nonlinearity (a condensate
with $a>0$ also oscillates without dispersion in the presence of harmonic
confinement\cite{Dobson94}). At each value of ${N/ N_{c}}$, the measured
axial density profiles were used to calculate the average harmonic potential energy
per atom at different times during the oscillation, and subsequently fit to
determine the oscillation frequency (see Methods). Oscillations for ${N/
N_{c}=-0.53}$, ${N/ N_{c}=0}$, and ${N/ N_{c}=+0.55}$ are plotted in Fig.~%
\ref{fig:fig4}b for a total of three trap periods, in each case. Since the
potential energy per atom is plotted, a total of six oscillations are
observed. The frequency for the attractive (repulsive) interactions clearly
leads (lags) when compared to the non-interacting ($a=0$) case. The relative
frequency shifts are plotted in Figure \ref{fig:fig4}c and we find them to
be in reasonable agreement with numerical simulations obtained by solving
the $1$D GPE. We observe that the relative shift also provides
a sensitive measurement of the zero-crossing, which is in excellent
agreement with a previous determination\cite{Pollack09}. The frequency shift
is independent of $\Delta\phi$, indicating that it is unrelated to the
phase-dependent interactions previously discussed.

We have developed a simple analytical approximation to calculate the frequency shift.  The shift arises from the mean-field interaction in which one soliton changes the potential landscape experienced by the other soliton. The phase shift is dominated by the
incoherent (density-density) terms in the interaction, and we neglect all other interaction terms in the GPE in comparison.  This approximation is valid for relatively weak nonlinearity and for fast moving solitons (see Methods).  Although a phase shift is expected in integrable systems, the analytical approximation accounts for harmonic confinement and thus provides an estimate of the frequency shift for either sign of the nonlinearity. The analytically predicted relative frequency shift is $\Delta \omega_{z} /
\omega_{z} = -0.67(N/N_{c}) a_{z}^{4}/\pi z_{0}^{3}a_{r}$, in which $z_{0}$
is the oscillation amplitude of a single soliton, and $a_{z}$ ($a_{r}$) is
the axial (radial) harmonic oscillator length.  $N / N_{c}$ is negative for $a < 0$. This approximation provides a simple,
intuitive picture. For ${a<0}$, the attraction between atoms accelerates the
solitons as they approach one another, and decelerates them as they move
away. The same occurs for ${a>0}$, but with opposite sign. A similar effect
has been observed for repulsive condensates in which oscillations of one
condensate induced motion in the other\cite{Maddaloni00, Modugno00}.

Our studies elucidate the role of integrability, relative phase, spatial
dimensionality, and mean-field interactions in soliton collisions. A natural
extension of this work would involve control over the relative phase between
solitons, and better control of the strength of the nonlinearity. This would enable us to study collisions in a more controlled manner,
providing the ability to further explore the transition between integrable
and non-integrable systems, and to study the formation of soliton molecules%
\cite{Khawaja11}. Finally, this geometry may be applicable to atom soliton
interferometry, demonstrated recently using a Bragg beamsplitter\cite%
{McDonald14} rather than the tunnel barrier adopted in our geometry.

\begin{methods}

%\section*{Methods}

\subsection{Apparatus}

The primary difference between the apparatus used here and that which was described previously\cite{Dries10} is that a pair of perpendicularly oriented laser beams, derived from a single fiber laser operating at ${1,\!070\ \mathrm{nm}}$, provide
cylindrically-symmetric harmonic confinement. The beam is divided
into two separate paths, directed parallel and perpendicular to the magnetic
field axis, and focused at the atoms to a ${1/e^{2}}$ radius of ${28\ \mathrm{\mu m}}$ and ${105\ \mathrm{\mu m}}$, respectively.

The magnetic field is controlled using a pair of coils in Helmholtz
configuration, and allows us to vary the scattering length across a broad
region. Initially, the BEC is formed at a field of ${716\ \mathrm{G}}$,
corresponding to a scattering length of ${a\approx 140a_{0}}$\cite{Dyke13}.
Once the BEC is formed, a blue-detuned Gaussian beam is turned on to cut the
condensate in half and act as a high barrier between the two halves. The field is
adiabatically ramped down (${t=750\ \mathrm{ms}}$) to a final scattering
length of ${a=-0.57\mathrm{a_{0}}}$.  We find experimentally that this procedure produces two solitons that interact with a randomly distributed relative phase.

To measure the oscillation frequencies for different nonlinearities, the final scattering length is varied between ${a=-0.57\mathrm{a_{0}}}$ and $a=+0.37\mathrm{a_{0}}$, corresponding to nonlinear strengths of $N/N_{c}=-0.53$ and ${N/N_{c}=+0.55}$, respectively.

The barrier is a cylindrically focused beam, blue-detuned by $900\ \mathrm{GHz}$
from the $2$S-$2$P resonance, with a ${1/e^{2}}$ radius of $2\ \mathrm{mm}$ perpendicular to the condensate, and a ${1/e^{2}}$ radius of $5.6\ \mathrm{\mu m}$ along the condensate axis. A barrier height of approximately $2\ \mathrm{\mu K}$ was used to
split the condensate, and to maintain a centre-to-centre separation of $26\
\mathrm{\mu m}$ between solitons.

Polarisation phase-contrast imaging (PPCI) was used in order to minimally
perturb the atoms, allowing us to take multiple images during a single
experimental run. Since the relative phase of solitons varies between
experimental runs, the use of this technique was crucial for
interpreting the collisional dynamics. PPCI exploits the birefringence of
the scattered light from atoms in a strong magnetic field. The scattered
light is interfered with the probe light using a linear polariser. The
resulting image is simply related to the column density distribution\cite{Bradley97}. With this technique the laser may be far detuned from resonance ($35\Gamma $ in this case, where $\Gamma =5.9\ \mathrm{MHz}$), minimizing
the number of photons scattered during the imaging process. Furthermore, the
$1/e^{2}$ beam radius of approximately $11\ \mathrm{mm}$ provides a uniform
intensity profile across the soliton pair so that any phase-shift imprinted on the solitons from
the imaging beam is common for the pair.

\subsection{Oscillation frequency}

The axial density, $n_{1D}(z,t)$, was calculated for each image and used to
determine the potential energy per atom from:

\begin{equation}
U(t)=\frac{1}{N \hbar \omega_{z}}{\int_{-\infty}^{\infty}n_{1D}(z,t)[\frac{1%
}{2} m (\omega_{z} z(t))^{2}]dz}.
\end{equation}

\subsection{Analytical approximation of frequency shift}

The quasi-1D GPE is:
\begin{equation}
i\hbar \frac{\partial \psi}{\partial t}=-\frac{\hbar ^{2}}{2m}\frac{\partial^{2} \psi}{\partial z^{2}}+\frac{1}{2}m\omega
_{z}^{2}z^{2}\psi +g_{1d}|\psi |^{2}\psi ,  \label{GPE}
\end{equation}%
in which $g_{1d}=2\hbar ^{2}a/ma_{r}^{2}$. Here, $a$, as above, is the
atomic scattering length, and $a_{z}(a_{r})$ is the axial (radial) harmonic
oscillator length\cite{Olshanii98}. In the absence of the nonlinear
interaction, the two-soliton state is modeled by $\psi =\psi _{1}+e^{i\phi
}\psi _{2}$, and $\psi _{i}$ is:
\begin{equation}
\psi _{i}=\left( \frac{m\omega _{z}N^{2}}{\pi \hbar }\right) ^{1/4}\exp
\left( -\frac{i}{2}\hbar \omega _{z}t-\frac{i}{2}m\omega _{z}^{2}\xi
_{i}^{2}+im\frac{d \xi_{i}}{dt} z\right) \exp \left( -\frac{m\omega _{z}}{2\hbar }%
(z-\xi _{i})^{2}\right) .  \label{psi}
\end{equation}%
We have introduced the position coordinate $\xi $ such that $\xi =\xi
_{1}=-\xi _{2}=z_{0}\sin (\omega _{z}t)$, which defines a pair of symmetric
Gaussians in the harmonic trap. In the limit of large impact speed (i.e. $%
z_{0}\gg a_{z}$) the interaction Hamiltonian becomes:
\begin{equation}
U_{\mathrm{int}}(\xi )=2g_{1d}\int_{-\infty }^{\infty }|\psi _{1}|^{2}|\psi
_{2}|^{2}dz,  \label{interaction
potential}
\end{equation}%
in which the coherent interaction terms are neglected due to the fast
spatial-phase oscillations between the rapidly moving solitons. We treat the
interaction-induced shift as a small perturbation, and write the soliton
motion as $\xi (t)=z_{0}\sin (\omega _{z}t)+\Delta \xi $. The equation of
motion for the perturbation $\Delta \xi $ is\cite{Malomed02}:
\begin{equation}
Nm\frac{d^{2}}{dt^{2}}\Delta \xi =-\frac{1}{2}\frac{d}{d\xi }U_{\mathrm{int}%
}(\xi ),  \label{EOM}
\end{equation}%
in which $Nm$ serves as the effective mass of the soliton and the factor $1/2
$ is due to the identity $\xi \equiv (\xi _{1}-\xi _{2})/2$. By substituting
Eqns.~(\ref{psi}) and (\ref{interaction potential}) in Eqn.~(\ref{EOM}), we
find that the total spatial jump, due to the density-density interaction, is:
\begin{equation}
\Delta \xi =-\frac{g_{1d}Na_{z}^{2}}{2z_{0}^{2}\hbar \omega _{z}},
\end{equation}%
and the corresponding shift in the oscillation frequency is:
\begin{equation}
\frac{\Delta \omega _{z}}{\omega _{z}}=-\frac{g_{1d}Na_{z}^{2}}{2\pi
z_{0}^{3}\hbar \omega _{z}}=-\frac{0.67(N/N_{c})a_{z}^{4}}{\pi z_{0}^{3}a_{r}%
}.
\end{equation}%
The approximate analytical approach presented here applies to
a broad class of pulses in generic models, matching the known results for integrable ones.

\subsection{Uncertainties}

The uncertainty in the strength of the nonlinearity is due to the
uncertainty in the atom number, $N$, the determination of the scattering
length, $a$, and the radial trap frequency, $\omega_{r}$. The uncertainty in
$N$ arises from $20\%$ shot-to-shot variation in $N$ and a systematic
uncertainty of $12 \%$ due to our ability to discern atoms from the
background. To measure $\omega_{r}$, the trap intensity was modulated near
the radial trap frequency and the resultant loss in atom number, from
heating, was measured. The uncertainty in $\omega_{r}$ determined from a
Lorentzian fit to the data is $<1 \%$. The mapping of $a$ vs. $B$ has been
previously determined, with our region of interest being near the
zero-crossing\cite{Pollack09}. A linear fit to the data near the zero
crossing gives a slope of ${0.08(1)\ a_{0}/\mathrm{G}}$ and a zero crossing
crossing at $B_{0}=543.6(1)\ \mathrm{G}$, with the uncertainties derived
from a systematic uncertainty in the field calibration of $0.1\ \mathrm{G}$.
This gives a systematic uncertainty in $a$ of $20 \%$ for $|a|\simeq 0.5a_{0}$. Thus, the statistical
and systematic uncertainties in the strength of the nonlinearity are $20\%$
and $23\%$, respectively, for $|a|\simeq 0.5a_{0}$.

\end{methods}

\begin{addendum}
 \item We thank Rodrigo Shiozaki for his help with the data acquisition system, and Maxim Olshanii
 for helpful discussions.  Support for this work was provided by the NSF, ONR, the Welch Foundation
 (Grant C-1133), the Binational (US-Israel) Science Foundation (Grant No. 2010239), and an ARO-MURI Non-equilibrium many-body dynamics grant (W911NF-14-1-0003).
 \item[Author contributions] J.N., P.D., and D.L. performed the experiment and analysed the data.
 D.L. performed the numerical simulations.  B.M. developed the analytical model.  R.H. was involved
 in all aspects of the experiment.  All authors discussed the results and implications and took part in
 preparing the manuscript.
 \item[Competing Interests] The authors declare that they have no competing financial interests.
 \item[Correspondence] Correspondence and requests for materials should be addressed
 to R.G.H.~(email: randy@rice.edu).
\end{addendum}

\newpage

\begin{figure}
\centering
\includegraphics{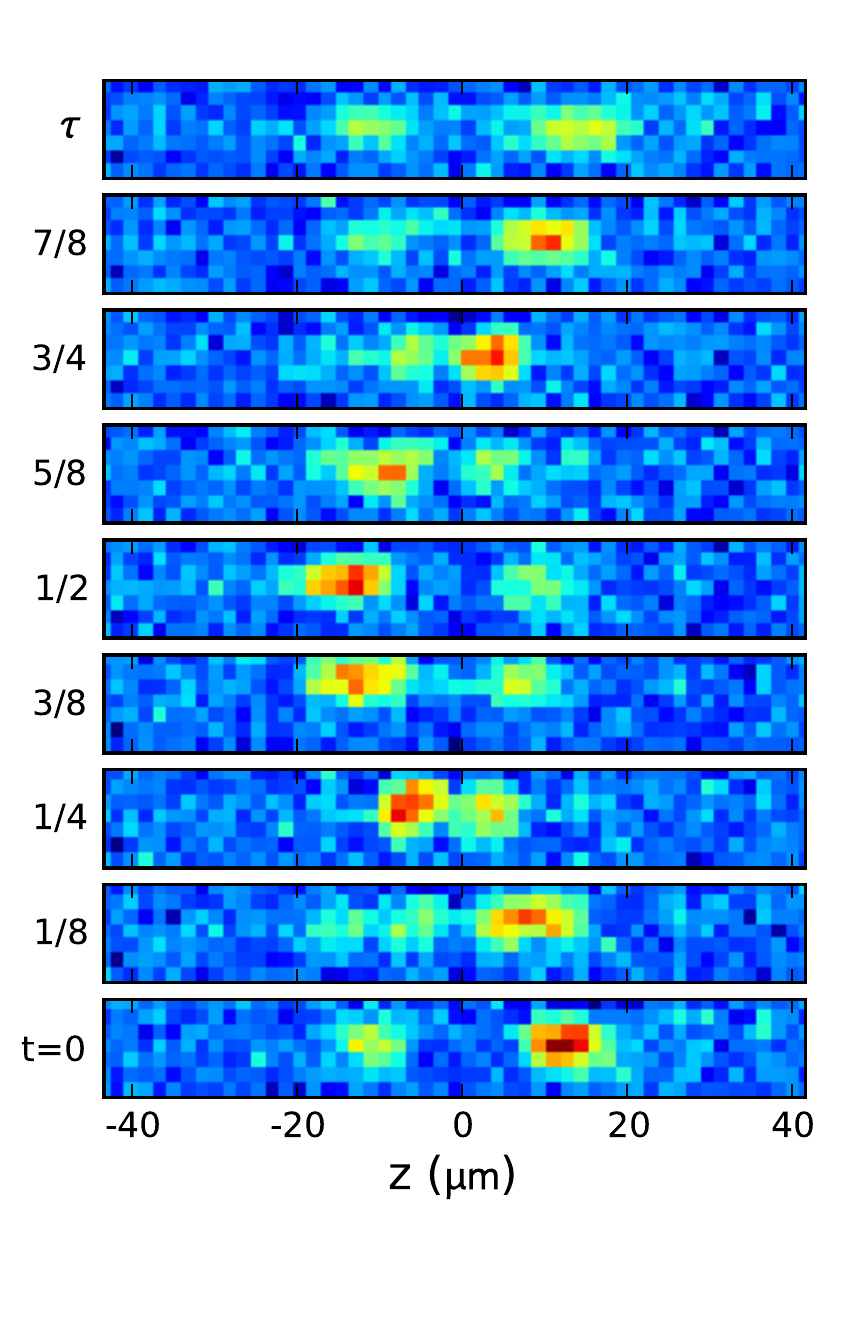}
\caption{\textbf{Tagged trajectory of soliton collision.} A pair of solitons
is formed with a ratio of ${2\!\!:\!\!1}$ in atom number. The resultant
collision appears to be repulsive, indicated by the density minimum
appearing between the pair at ${t=1/4\protect\tau}$ and ${3/4\protect\tau}$.
However, by following the trajectories of individual solitons, we observe
that they actually pass through one another in the course of each collision.}
\label{fig:fig3}
\end{figure}

\begin{figure}
\centering
\includegraphics{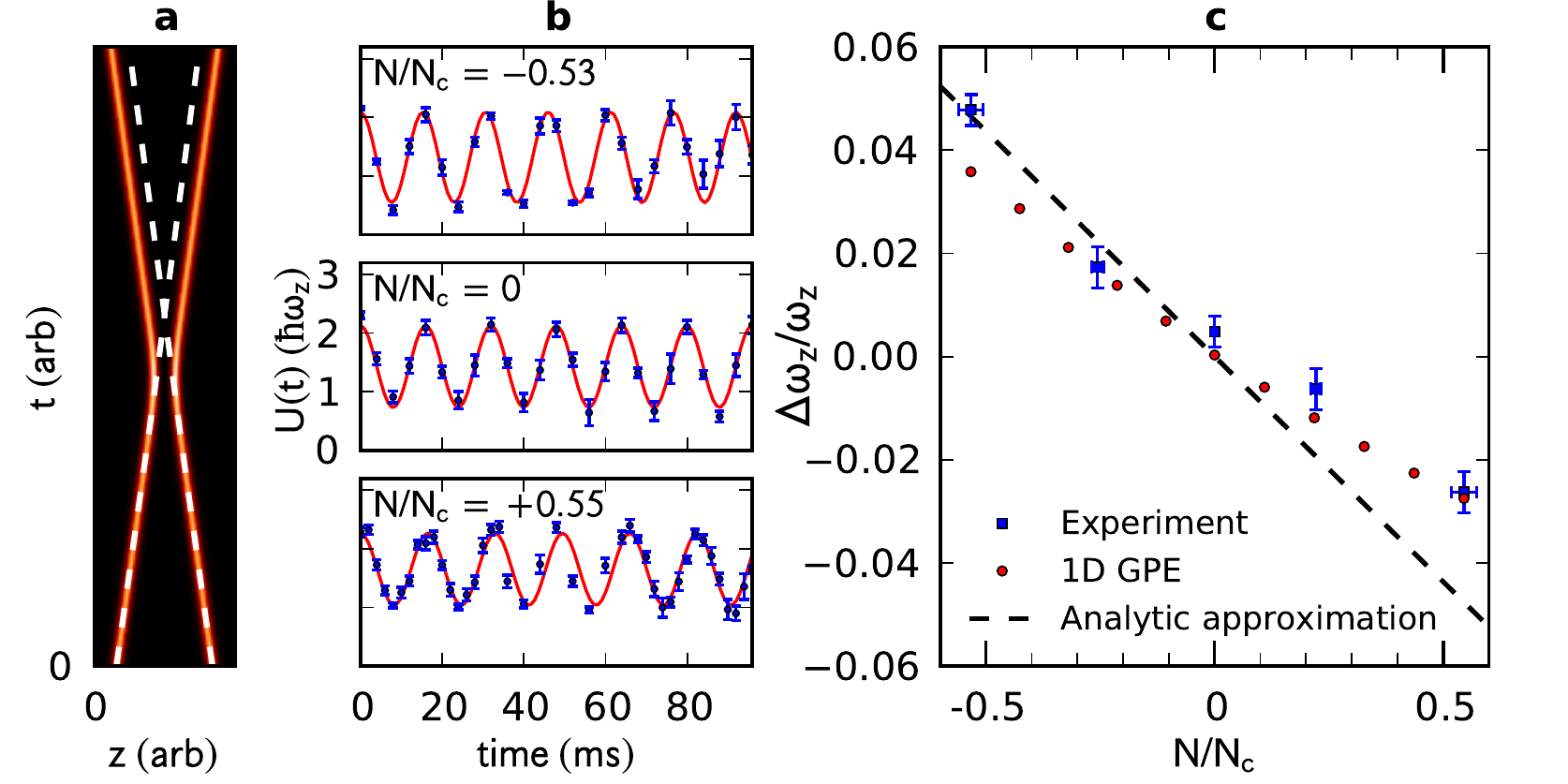}
\caption{\textbf{Frequency shift due to mean-field interaction between
solitons.} \textbf{a}, Simulated trajectory without an axial potential, in
dimensionless units. The dashed-white lines shows the soliton trajectories
in the absence of interaction. \textbf{b}, The harmonic potential energy per
atom, $U(t)$ (see Methods), is plotted for different strengths of the
nonlinearity. Each data point is the mean of five different experimental
runs (blue points).  The red lines are the results of a fit to determine the oscillation
frequency. The error bars correspond to the standard error of the mean. \textbf{c}, The
relative frequency shifts determined from the fits to the experimental data
are plotted vs. the nonlinearity strength (blue squares). The error bars
correspond to the standard error of the fit for $\Delta\protect\omega_{z}/ \protect\omega%
_{z}$. The error bars for $N/N_{c}$ are the standard error of the mean for $5$
measurements of $N$. In addition, the systematic uncertainty in $N/N_{c}$ is
estimated to be $23\%$ for $|a|\simeq 0.5a_{0}$ (see Methods). Relative shifts were also determined
by numerically solving the $1$D-GPE (red points). An analytical
approximation determined solely by the incoherent density-density terms in
the GPE is shown by the dashed line, as described in Methods.}
\label{fig:fig4}
\end{figure}

\begin{figure}[h!b]
\centering
\includegraphics{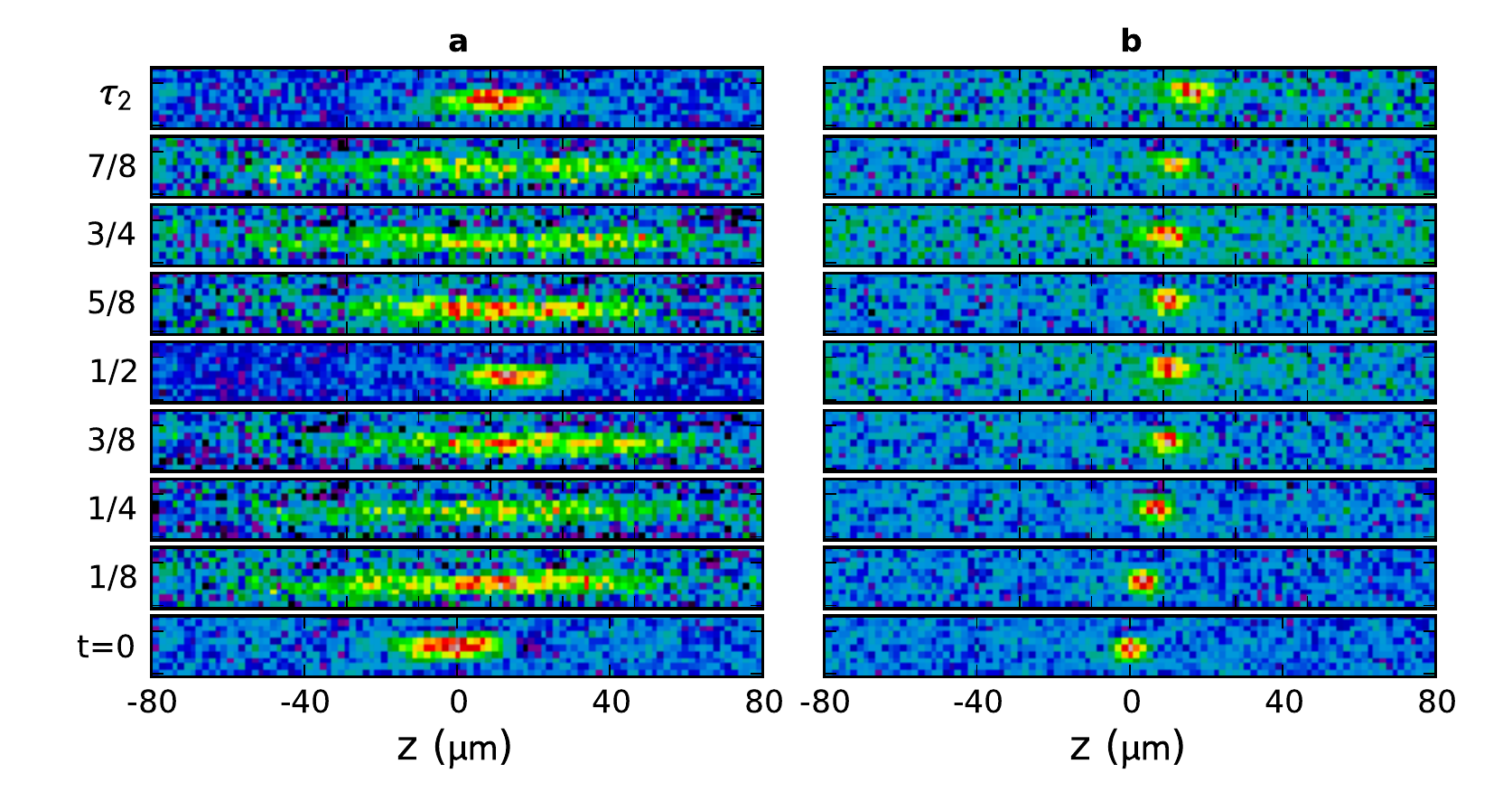}
\caption*{Supplementary Figure 1: \textbf{Comparison between the expansion of
a soliton and a repulsive condensate.} \textbf{a}, A condensate with
repulsive interactions is formed with ${N/ N_{c}=+0.55}$, and suddenly
transferred into a single-beam trap by turning off the beam perpendicular
to the magnetic field axis. The new axial trap frequency is ${\protect\omega%
_{z2}= 8\ \mathrm{Hz}}$ (${\protect\tau_{2}=125\ \mathrm{ms}}$). Images are
taken at intervals of ${\tau_{2} / 8}$.  The condensate rapidly expands as a
result of the repulsive interactions between atoms. The subsequent
reformation of the condensate at ${\tau_{2} / 2}$ and at ${\tau_{2}}$ indicates that a breathing mode has been excited. \textbf{b}, Similar to \textbf{a}, but with ${N/
N_{c}=-0.53}$. Here, the soliton propagates without dispersion, as expected.
The small displacement in time is due to a slight time-dependent
displacement between trap centres.  Note that the color mappings of Supplementary Figs. $1$a and $1$b differ to account for differences in density.}
\label{fig:supfig}
\end{figure}

\end{document}